\documentstyle[psfig]{mn}
\newif\ifAMStwofonts

\newcommand{\feh}{\hbox{$[{\rm Fe}/{\rm H}]$}}
\newcommand{\mvrr}{\hbox {${\rm M_v(RR)}$}}
\newcommand{\mvto}{\hbox {${\rm M_v(TO)}$}}
\newcommand{\mvb}{\hbox {${\rm M_v(BTO)}$}}
\newcommand{\vb}{\hbox {${\rm V(BTO)}$}}
\newcommand{\dv}{\hbox {$\Delta \rm V(TO-HB)$}}
\newcommand{\dvb}{\hbox {$\Delta \rm V(BTO-HB)$}}
\newcommand{\dbv}{\hbox {$\Delta \rm (B-V)$}}
\newcommand{\ea}{{\it et al.}}

\ifoldfss
  \ifCUPmtlplainloaded \else
    \NewTextAlphabet{textbfit} {cmbxti10} {}
    \NewTextAlphabet{textbfss} {cmssbx10} {}
    \NewMathAlphabet{mathbfit} {cmbxti10} {} 
    \NewMathAlphabet{mathbfss} {cmssbx10} {} 
  \fi
  \ifAMStwofonts
    \ifCUPmtlplainloaded \else
      \NewSymbolFont{upmath} {eurm10}
      \NewSymbolFont{AMSa} {msam10}
      \NewMathSymbol{\upi}     {0}{upmath}{19}
      \NewMathSymbol{\umu}     {0}{upmath}{16}
      \NewMathSymbol{\upartial}{0}{upmath}{40}
      \NewMathSymbol{\leqslant}{3}{AMSa}{36}
      \NewMathSymbol{\geqslant}{3}{AMSa}{3E}

       \let\le=\leqslant
       
    \fi
  \fi
\fi 

\ifnfssone
  \newmathalphabet{\mathit}
  \addtoversion{normal}{\mathit}{cmr}{m}{it}
  \addtoversion{bold}{\mathit}{cmr}{bx}{it}
  \newmathalphabet{\mathbfit} 
  \addtoversion{normal}{\mathbfit}{cmr}{bx}{it}
  \addtoversion{bold}{\mathbfit}{cmr}{bx}{it}
  \newmathalphabet{\mathbfss} 
  \addtoversion{normal}{\mathbfss}{cmss}{bx}{n}
  \addtoversion{bold}{\mathbfss}{cmss}{bx}{n}
  \ifAMStwofonts
    \ifCUPmtlplainloaded \else
      \UseAMStwoboldmath
      \makeatletter
      \new@mathgroup\upmath@group
      \define@mathgroup\mv@normal\upmath@group{eur}{m}{n}
      \define@mathgroup\mv@bold\upmath@group{eur}{b}{n}
      \edef\UPM{\hexnumber\upmath@group}
      \new@mathgroup\amsa@group
      \define@mathgroup\mv@normal\amsa@group{msa}{m}{n}
      \define@mathgroup\mv@bold\amsa@group{msa}{m}{n}
      \edef\AMSa{\hexnumber\amsa@group}
      \makeatother
      \mathchardef\upi="0\UPM19
      \mathchardef\umu="0\UPM16
      \mathchardef\upartial="0\UPM40
      \mathchardef\leqslant="3\AMSa36
      \mathchardef\geqslant="3\AMSa3E

       \let\le=\leqslant

    \fi
  \fi
\fi 

\ifnfsstwo
  \DeclareMathAlphabet{\mathbfit}{OT1}{cmr}{bx}{it}
  \SetMathAlphabet\mathbfit{bold}{OT1}{cmr}{bx}{it}
  \DeclareMathAlphabet{\mathbfss}{OT1}{cmss}{bx}{n}
  \SetMathAlphabet\mathbfss{bold}{OT1}{cmss}{bx}{n}
  \ifAMStwofonts
    \ifCUPmtlplainloaded \else
      \DeclareSymbolFont{UPM}{U}{eur}{m}{n}
      \SetSymbolFont{UPM}{bold}{U}{eur}{b}{n}
      \DeclareSymbolFont{AMSa}{U}{msa}{m}{n}
      \DeclareMathSymbol{\upi}{0}{UPM}{"19}
      \DeclareMathSymbol{\umu}{0}{UPM}{"16}
      \DeclareMathSymbol{\upartial}{0}{UPM}{"40}
      \DeclareMathSymbol{\leqslant}{3}{AMSa}{"36}
      \DeclareMathSymbol{\geqslant}{3}{AMSa}{"3E}

       \let\le=\leqslant

    \fi
  \fi
\fi 

\ifCUPmtlplainloaded \else
  \ifAMStwofonts \else 
    \def\upi{\pi}
    \def\umu{\mu}
    \def\upartial{\partial}
  \fi
\fi

\title[A Precision Age Estimator]{A Precision Age Determination \qquad
\qquad \raisebox{20pt}[0pt]{\large CITA-96-6 ~~~CWRU-P6-96}\\
Technique for Globular Clusters}

\author[B. Chaboyer et al.]
{Brian Chaboyer,$^1$ Pierre Demarque,$^2$ Peter J.\ Kernan,$^3$
\newauthor 
Lawrence M.\ Krauss$^3$\thanks{Also Dept of Astronomy} and
Ata Sarajedini$^4$\thanks{Hubble Fellow}\\
$^1$CITA, 60 St.\ George St., Toronto, ON, Canada M5S 3H8~~E-Mail:
chaboyer@cita.utoronto.ca \\
$^2$Department of Astronomy, Yale University, New Haven, CT, USA~~
06520-8101~~E-Mail: demarque@astro.yale.edu\\
$^3$Department of Physics, Case Western Reserve University, 10900
Euclid Ave., Cleveland, OH, USA~~ 44106-7079\\~~E-Mail:
pete@theory2.phys.cwru.edu; krauss@theory1.phys.cwru.edu\\
$^4$Kitt Peak National Observatory, National Optical Astronomy
Observatories\thanks{NOAO is operated by the Association of
Universities for Research in Astronomy, Inc., under contract with the
National Science Foundation.}, PO Box 26732, Tucson, AZ, USA~~ 85726\\
~~E-Mail:ata@noao.edu}

\date{\raisebox{10pt}[0pt]{Submitted April 11/96}}
\pagerange{\pageref{firstpage}--\pageref{lastpage}}
\pubyear{1996}

\def\LaTeX{L\kern-.36em\raise.3ex\hbox{a}\kern-.15em
    T\kern-.1667em\lower.7ex\hbox{E}\kern-.125emX}

\begin{document}

\label{firstpage}

\maketitle

\begin{abstract}
Globular cluster age estimates based on the absolute magnitude of the
main sequence turn-off (\mvto) are generally considered to be the most
reliable from a theoretical viewpoint.  However, the difficulty in
determining \mvto\ in observed colour-magnitude diagrams leads to a
large error in the derived age.  In this paper, we advocate the use of
the absolute magnitude of the point which is brighter than the
turn-off and 0.05 mag redder (\mvb) as a precision age indicator.  It
is easy to measure this point on observed colour-magnitude diagrams,
leading to small observational error bars.  Furthermore, an extensive
Monte Carlo calculation indicates that the theoretical uncertainty in
\mvb\ is similar to \mvto.  As a result, ages derived using \mvb\ are
at least a factor of 2 more precise than those derived using \mvto.
This technique is applied to the globular cluster M68 and an age of
$12.8\pm 0.3\,$Gyr is derived (assuming $\mvrr = 0.20\,\feh +
0.98$), indicating that M68 is a `young' globular cluster.  A
homogeneous set of globular cluster age estimates with this precision
would provide unprecedented insight into the formation of the Galactic
halo.

\end{abstract}



\section{Introduction}

There are a number of different techniques which may be used to
determine the age of a globular cluster (GC). All of these methods
rely on comparing some aspect of theoretical stellar evolution models
to the observations.  Thus, in order to evaluate the reliability of
the various age indicators, one must be aware of the uncertainties in
theoretical stellar evolution models.  The correct treatment of
convection in stellar models is an area of active research (e.g.\ Kim
\ea\ 1995, 1996; Demarque, Guenther \& Kim 1996a,b) and remains
the largest possible source of error in stellar models.  For this
reason, properties of the stellar models which depend on the treatment
of convection are the most uncertain.  The main sequence and red giant
branch stars in GCs have surface convection zones, and so the
predicted radii (and hence, colours) are subject to large theoretical
uncertainties.  The helium burning stars (horizontal branch, and
asymptotic giant branches) are convective in the energy generation
regions, and so even the predicted lifetimes and luminosities of stars
in this phase of evolution are somewhat uncertain.  An additional
consideration when considering the reliability of stellar models is
that observed CNO abundances in stars on the red giant branch indicate
that some form of deep mixing occurs in these stars, which is {\em
not} present in the models (e.g.\ Langer et al.\ 1983; Kraft 1994;
Chaboyer 1995).  This indicates that the red giant branch models are
in need of revision.  In contrast, low mass main sequence models are
in excellent agreement with the observations. Indeed, inversions of
solar models which use the observed $p$-modes indicate that the run of
density and sound speed in solar models agree with the Sun to within
1\% \cite{basu}.

The relative reliability of the age-luminosity relationship for low
mass stars is well known, and it is for this reason that the absolute
magnitude of main sequence turn-off (\mvto) results in GC ages with
the smallest theoretical error (e.g.\ Renzini 1991).  Operationally,
\mvto\ is defined to be the magnitude of the bluest point on the main
sequence.  (Since this definition involves the use of colour \mvto\ is
not strictly independent of the uncertainties in stellar radii.)
Unfortunately, the turn-off region has nearly the same colour over a
large range in magnitude.  This leads to difficulties measuring \mvto\
observationally, due to the scatter in the observed points around the
turn-off.  Observers typically quote errors of order 0.10 mag in
\mvto, which leads to an error in the derived age around $\pm
1.5\,$Gyr (e.g.\ (Sarajedini \& King 1989; Chaboyer, Demarque \& Sarajedini 1996, hereafter
CDS).  This large error in the derived age of any individual GC is a
great obstacle in furthering our understanding of galaxy formation.
This problem has lead Sarajedini \& Demarque (1990) and VandenBerg,
Bolte \& Stetson (1990) to advocate the use of the difference in
colour between the main sequence turn-off and the base of the giant
branch (\dbv) as an age indicator.  This method has the advantage that
the colour of the turn-off and the base of the giant branch can be
accurately determined in observed colour-magnitude diagrams (CMDs),
and is independent of the distance modulus.  As a result, this method
can lead {\it in principle\/} to very precise age estimates (of order
$\pm 0.5\,$Gyr).  However, the theoretical colours are subject to
large uncertainties and \dbv\ only yields reliable {\em relative\/}
age differences between clusters of a similar metallicity (see,
however, the case of M68 and M92 (\S \ref{sec3}) for which \dbv\
fails).

In this paper, we advocate the use of a point which is brighter than
the turn-off, and 0.05 mag redder in B--V (hereafter referred to as
\mvb).  This point is easy to measure in observational data and has a
small theoretical uncertainty.  As it still requires  knowledge of
the distance modulus, \mvb\ complements the \dbv\ technique in
providing precision age estimates for GCs.  In \S \ref{sec2}, a Monte
Carlo set of isochrones is described and analyzed in order to
estimate the theoretical error associated with \mvb\ and \mvto.  The
well studied GC M68 is used to illustrate the relative precision of
ages derived using \mvb\ and \mvto\ in \S \ref{sec3}.  Finally, \S
\ref{sec4} summarizes the results of this work and suggests that
observers should quote \mvb\ in their papers in addition to \mvto.
Simple formulae are provided to determine GC ages, given \mvb\ and
\feh.

\section{Theoretical Analysis}\label{sec2}
The basic problem in measuring \mvto\ is that the turn-off region
is nearly vertical in the HR diagram.  Thus, the colour of the main
sequence turn-off is well defined, but its magnitude is not.  As stars
evolve off the main sequence they  quickly expand, and so
points somewhat brighter than the turn-off are more horizontal in the
HR diagram.  Thus, it is easy to measure the magnitude of \mvb, and
ages derived using \mvb\ will have small observational error
bars.  The main reason for using \mvto\ as an age indicator is that it
is widely perceived to be the most robust of the theoretical age
estimators. Thus, ages derived using \mvto\ have small
theoretical `error bars'.  The key question then is whether the
theoretical uncertainty in \mvb\ is similar to that in \mvto.  If that
is the case, then there would be no reason to use \mvto\ in GC age
estimates.  

The theoretical error in \mvb\ may be estimated by constructing a
series of isochrones under a variety of assumptions.  In a study
designed to provide an estimate of the error associated with GC ages
Chaboyer, Demarque, Kernan \& Krauss (1996, hereafter CDKK) calculated
1000 independent sets of isochrones.  These isochrones were
constructed via a Monte Carlo analysis, whereby the various input
elements needed to compute a stellar model and isochrone (such as
opacity, mixing length, etc.)  were picked at random from
distributions based on a careful analysis of the recent literature.
Table \ref{tab1} provides an outline of the various input parameters
and their distribution.  Further details are provided in CDKK.  This
Monte Carlo study was designed to yield a set of isochrones which span
the range of relevant uncertainties in modern stellar evolution
calculations.
\begin{table*}
\begin{minipage}{125mm}
\caption{Monte Carlo Input Parameters}\label{tab1}
\begin{tabular}{lll}
\hline\hline
\multicolumn{1}{c}{Parameter}&
\multicolumn{1}{c}{Distribution}&
\multicolumn{1}{c}{Comment}\\[2pt]
\hline
mixing length & $1.85\pm 0.25$ (stat.) & fits GC observations\\
helium diffusion coefficients & 0.3 -- 1.2 (syst.) & possible
systematic error dominate\\
nuclear reaction rates & \multicolumn{2}{l}{see CDKK}\\
OPAL high temperature & $1\pm 0.01$ (stat.) & comparison of
OPAL\\
~~~opacities & & \& LAOL opacities\\
surface boundary condition & \multicolumn{2}{l}{grey or Krishna-Swamy (1966)}\\
colour table & \multicolumn{2}{l}{Green \ea\ (1987) or Kurucz (1992)}
\\
primordial $^4$He abundance & $0.22 - 0.25$ (syst.)&possible
systematic error dominate\\
oxygen abundance, [O/Fe] & $+0.55\pm 0.05$ (stat.)\\
& $\pm 0.20$(syst.)\\
\hline
\end{tabular}
\end{minipage}
\end{table*}

In the original work, each set of isochrones consisted of 45
isochrones in the age range 8 -- 22 Gyr, with metallicities $\feh
=-2.5,$ $-2.0$ and $-1.5$.  We have supplemented this with additional
calculations with $\feh = -1.0$ and $-0.5$ to span the majority of the
GC metallicity range.  The three lowest metallicity isochrones in the
original set assumed that the helium abundance was equal to its
primordial value ($Y_P$).  The higher metallicity isochrones in the
new calculations allowed for some helium enrichment, assumed to be $Y
= Y_P + 1.8\, \Delta Z$, where $Z$ is the mass fraction of heavy
elements. In addition, the $\feh = -0.5$ isochrone assumed that the
oxygen enhancement was one-half (in dex) of that in the more metal-poor
isochrones.  In total, about $5\times 10^4$ stellar evolutionary runs
were performed, involving the calculations of nearly 8 million stellar
models.

The Monte Carlo set of 14 Gyr, $\feh = -2.0$ isochrones is shown in
Figure 1, with the turn-off and brighter points highlighted.
This figure demonstrates the very wide range in colour which is
possible in theoretical isochrones, given the uncertainties in present
stellar models and isochrone construction.  As expected, the range
\mvto\ is rather small.  The 1-$\sigma$ (68\% confidence limits) range
in \mvto\ is $\pm 0.0620\,$mag.  The 1-$\sigma$ range in \mvb\ is
nearly identical, $\pm 0.0625\,$mag.  This analysis has been repeated
for the other metallicities, and in all cases the spread in \mvb\ was
found to be 
quite similar to the spread in \mvto.  The mean 1-$\sigma$ range was
$\pm 0.066\,$mag in \mvto, and $\pm 0.068\,$mag in \mvb.  This
analysis was repeated on a subset of 400 Monte Carlo isochrones with
ages of 10 and 18 Gyr and similar results were obtained.  This strongly
suggests that the theoretical error associated with ages derived
from \mvb\ will be similar to those derived from \mvto.
\begin{figure} \label{fig1}
\centerline{\psfig{figure=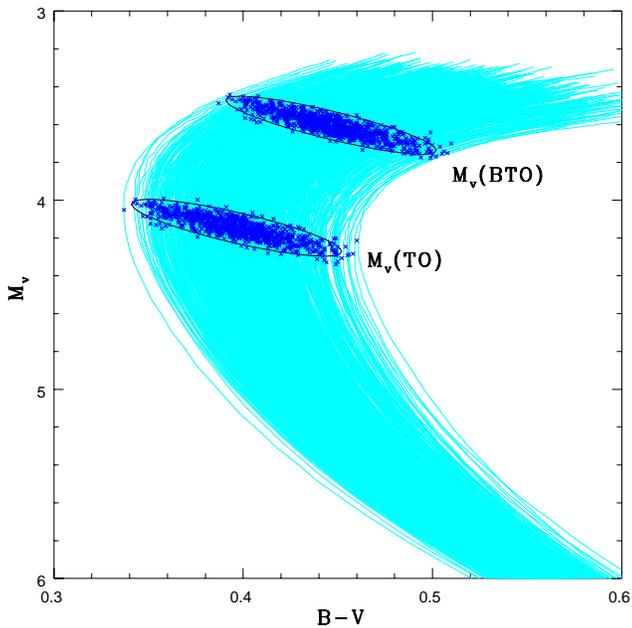,height=9.0cm}}
\caption{The turn-off region in the Monte Carlo set of 14 Gyr, $\feh =
-2.0$ isochrones. The turn-off (${\rm M_V} \sim 4.2$) and \mvb\ (${\rm
M_V} \sim 3.6$) points have been highlighted.}
\end{figure}

This issue may be addressed directly by comparing the spread in
derived ages for a given value of \mvto\ and \mvb.  Using the standard
set of isochrones described in the next paragraph, values of \mvto\
and \mvb\ were chosen which yielded an age of 15 Gyr.  These fixed
values of \mvto\ and \mvb\ (along with the corresponding metallicity)
were used as input parameters for a program which determined the
corresponding ages in each of our 1000 independent sets of isochrones.
The resulting set of 1000 \mvto\ and \mvb\ ages were analyzed.  The
dispersion in age was only slightly larger for ages derived using \mvb\ as
compared to \mvto.  For example, the 1-$\sigma$ dispersion at $\feh =
-2.0$ was $\pm 0.9\,$Gyr for the \mvto\ ages versus $\pm 1.0\,$Gyr for
the ages derived using \mvb.  This  indicates that the theoretical
uncertainty in ages derived using \mvb\ is similar to the
theoretical uncertainty in ages derived using \mvto.

Given that \mvb\ has a similar theoretical uncertainty to \mvto\ the
next important issue to address is the sensitivity of \mvb\ to age
changes. In order to evaluate this issue, a single set of isochrones
was constructed, with our best estimate for the input
physics\footnote{High temperature opacities from Iglesias \& Rogers
(1991); low temperature opacities from Kurucz (1991); nuclear reaction
rates from Bahcall \& Pinsonneault (1992) and Bahcall (1989); an
equation of state which includes the effects of Coulomb interactions
\cite{guenth,yc}; helium diffusion coefficients from Michaud \&
Proffitt (1993) multiplied by 0.75; a grey model atmosphere was used
for the surface boundary conditions, a near solar mixing length of
$\alpha = 1.85$ and the colour transformation of Green, Demarque \&
King (1987) was used to transform the isochrones to the observational
plane.} and composition.  For the primordial helium abundance, a value
of $Y=0.235$ was chosen.  The effect of the enhancement of the
$\alpha$-capture elements (O, Mg, Si, S, and Ca) was taken into
account by modifying the relationship between $Z$ and \feh, as
prescribed by \cite{scs}.  Over the range $2.5\le \feh \le -1.0$ a
value of $[\alpha/{\rm Fe}] = +0.55$ was employed \cite{nissen}, while
at $\feh = -0.5$, $[\alpha/{\rm Fe}] = +0.275$ was assumed.  In this
set of isochrones, the points \mvto\ and \mvb\ were determined and fit
to a simple quadratic of the form
\begin{equation}
t_9= a_o + a_1\,{\rm M_V} + a_2\,{\rm M_V}^2,  \label{fit}
\end{equation}
where $t_9$ is the age in units of Gyr.  The coefficients of the fit
at each metallicity are given in Table \ref{tab2}.  These coefficients
are valid for derived ages in the range 8 -- 22 Gyr.  
\begin{table*}
\begin{minipage}{105mm}
\caption{\mvto\ and \mvb\ Fit Coefficients (see eq.\  1)}
\label{tab2}
\begin{tabular}{lrrrcrrr}
\hline\hline
& \multicolumn{3}{c}{\mvto}& &
\multicolumn{3}{c}{\mvb}\\
\cline{2-4}\cline{6-8}
\multicolumn{1}{c}{\feh}&
\multicolumn{1}{c}{$a_o$}&
\multicolumn{1}{c}{$a_1$}&
\multicolumn{1}{c}{$a_2$}&&
\multicolumn{1}{c}{$a_o$}&
\multicolumn{1}{c}{$a_1$}&
\multicolumn{1}{c}{$a_2$}\\[2pt]
\hline
$-2.5$ & 2.261 & 0.1641 & $-0.003048$ && 1.886 & 0.1452 &$-0.002653$\\
$-2.0$ & 2.513 & 0.1506 & $-0.002696$ && 1.996 & 0.1521 &$-0.002866$\\
$-1.5$ & 2.596 & 0.1626 & $-0.003157$ && 2.190 & 0.1546 &$-0.002913$\\
$-1.0$ & 2.981 & 0.1343 & $-0.002353$ && 2.449 & 0.1503 &$-0.002654$\\
$-0.5$ & 3.063 & 0.1288 & $-0.002046$ && 2.563 & 0.1529 &$-0.002645$\\
\hline
\end{tabular}
\end{minipage}
\end{table*}

Figure 2 plots a subset of the standard set of isochrones for ages of
10, 14, 18 Gyr, and $\feh = -2.0$ (the same metallicity as in Fig.\
2).  This figure graphically illustrates that \mvto\ and
\mvb\ have similar sensitivity to age changes.  The sensitivity of
\mvto\ and \mvb\ as age indicators may be evaluated analytically by
taking the derivative of eq.\ 1.  A good age indicator will
have a large derivative, as this implies a small change in ${\rm M_V}$
results in a large change in age.  The ratio of the derivatives in
eq.\ 1 between \mvb\ to \mvto\ was calculated for all 5
metallicities and found to vary between 0.9 to 1.2.  This indicates
that \mvb\ has a similar sensitivity to age changes as \mvto\ over the
entire metallicity range tested.  Given that \mvb\ has a similar
theoretical uncertainty to \mvto, and that both are equally sensitive
to age changes, and that it is easier to measure
\mvb\ in observational databases, \mvb\ is clearly a superior age
diagnostic.
\begin{figure} \label{fig2}
\centerline{\psfig{figure=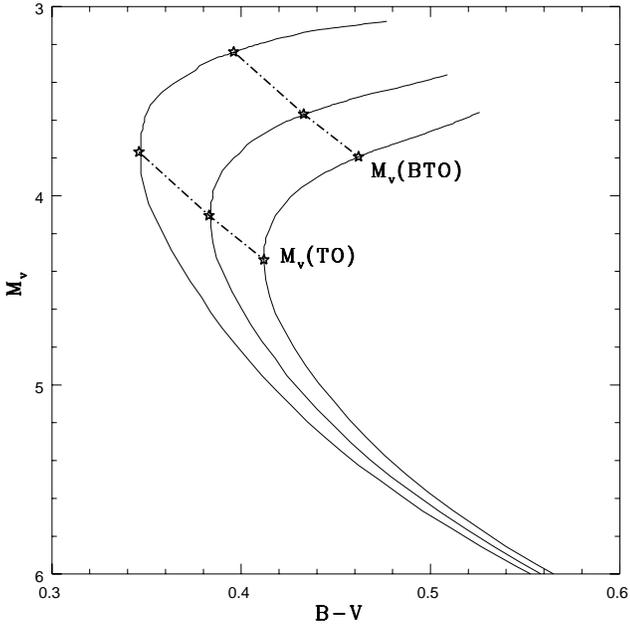,height=9.0cm}}
\caption{A sample of the $\feh = -2.0$  standard set of isochrones, 
with ages of 10, 14 and 18 Gyr.  The turn-off and \mvb\ points have
been highlighted.  Note that both points have a similar sensitivity to
age changes.}
\end{figure}

\section{Application to Observations}\label{sec3}
In order to determine ages using either \mvto\ or \mvb, the distance
modulus to the cluster must be known.  There are two main
techniques which may be used to determine the distance to a GC: (1) main
sequence fitting to local sub-dwarfs with well measured parallaxes, and
(2) using the observed magnitude of the horizontal branch (HB) combined with a
relationship for the absolute magnitude of the HB (derived using RR
Lyrae stars).  Unfortunately, there is only one sub-dwarf with a well
measured parallax (error in the absolute magnitude less than 0.05
mag), Groombridge 1830
\cite{ypc},
so the application of main sequence fitting to GCs is still rather uncertain.
Until improved sub-dwarf parallaxes become available, the
use of the HB to set the distance scale to globular clusters will
remain popular.  The HB has the advantage that the difference in
magnitude between the main-sequence turn-off and the horizontal branch
(\dv) is independent of reddening.  Thus, \dv\ is a widely used age
determination technique, which uses \mvto\ as its age diagnostic (e.g.\
CDS).  Although there are significant uncertainties in the
absolute magnitude of the RR Lyrae stars (used to determine \dv\ in
the theoretical calibration), CDS have shown that many statements can
be made regarding the {\em relative} GC ages which are independent of
the RR Lyrae calibration. This is very important, as the error 
in the absolute ages is dominated by the error in the distance
modulus (CDKK).  To exploit the merits of \mvb\ as an age indicator on
observations, we suggest the use of a modified $\Delta\,{\rm V}$
approach, using the difference in magnitude between \mvb\ and the HB,
\dvb\ in order to study relative GC ages, and the formation of the
Galactic halo.

The well studied GC M68 \cite{walker} provides an ideal database to
test this new age determination technique.  This is a relatively
metal-poor cluster with $\feh = -2.09\pm 0.11$ \cite{zinnw}.  Its HB
morphology is predominantly blue but significantly redder than other
clusters with comparable metallicity.  This led Zinn (1993) to
classify M68 as a relatively young halo cluster (see also Da Costa \&
Armandroff 1995). From Table 5 of Walker (1994), the mean V magnitude
of the RR Lyraes is $<{\rm V(RR)}> = 15.635\pm 0.006$ , where the
uncertainty is the standard error of the mean. There is also an error
in the photometric zero-point, but that is not included here because
$<{\rm V(RR)}>$ will be combined with V(TO) making the error in the absolute
photometric scale irrelevant.  For the theoretical HB magnitude, the
preferred relationship of CDS and CDKK, $\mvrr = 0.20\, \feh + 0.98$
was arrived at after a review of the current literature.  Recent HST
data \cite{ajhar} indicates that the slope is shallow, suggesting that
$\mvrr = 0.15\,\feh + 0.885$ may be a somewhat better choice.  However, as
emphasized by CDS, most statements regarding relative GC ages are
independent of the particular choice of \mvrr.

Walker (1994) measured a turn-off magnitude of ${\rm V(TO)} = 19.05\pm
0.05$, and hence $\dv = 3.415\pm 0.05$.  Due to the difficulty in
determining the turn-off point, Walker (1994) elected to increase his
error in \dv\ to $\pm 0.10\,$mag.  Using our standard set of
isochrones and $\mvrr = 0.20\, \feh + 0.98$ the \dv\ age of M68 is
$12.7\pm 1.3\,$Gyr using the larger error, and $12.7\pm 0.7\,$Gyr with
the smaller error bar in \dv.  If the shallower slope for \mvrr\ is
chosen ($\mvrr = 0.15\,\feh + 0.885$), an age of $12.8\pm 1.3$ is
derived. This age is 18\% younger than that derived by CDS, using the
same \mvrr\ relation.  The difference in age is due to the different
input physics used to construct the isochrones.  CDS ignored the
effects of diffusion (7\% increase in age), the Coloumb
correction to the equation of state (7\%) and assumed a somewhat
smaller $\alpha$-element enhancement ($[\alpha/{\rm Fe}] = +0.40$, 4\%).

In order to derive the age using \dvb, an objective technique was used
to measure \vb.  A fifth order polynomial (${\rm V} = f({\rm B-V})$)
was fitted to the stars in the region of the turnoff using a
2-$\sigma$ rejection scheme.  This polynomial yields $({\rm B-V})_{\rm
TO}$ and \vb.  The error in V(BTO) is determined by constructing a
histogram of the V deviations from the fit using stars within $\pm
0.03$\,mag in (${\rm B - V}$) of the \vb\ point.  A gaussian is then
fitted to the histogram. The standard deviation of this gaussian
divided by the square root of the number of data points used 
is then the 1-$\sigma$ error in \vb. Application of this
technique to the M68 data \cite{walker} results in $\vb = 18.519\pm
0.006$.  Note that the error in \vb\ is an order of magnitude smaller
than the error in V(TO).  

Combining the above value of \vb\ with the mean RR Lyrae magnitude
results in $\dvb = 2.884 \pm 0.008$ which implies an age of $12.8\pm
0.3\,$Gyr.  Due to the extremely high precision in the determination
of \dvb\ the error in the age is dominated by the error in metallicity
($\pm 0.11$ dex), and not by the error in measuring  \dvb.  If the
error in \dvb\ is increased to $\pm 0.025\,$mag, then the error in the
age increases slightly to $\pm 0.4\,$Gyr. An error of order $\pm
0.025$ in \dvb\ allows for a greater error  in the determination
of the HB level, and is perhaps more typical of most data in the
literature.  This example clearly demonstrates that ages derived using
\dvb\ are at least a factor of two more precise those derived using \dv.

CDKK constructed a sample of 17 metal-poor GCs (mean $\feh = -1.9$),
which were believed to be old based on \dbv\ or the horizontal branch
morphology.  The mean age of these 17 GC using our standard set of
isochrones is $15.2\pm 0.4\,$Gyr. Note that the  error in the \dvb\
age of M68 is smaller than the error in the mean age of 17 GCs
determined using \dv.  The difference in age between M68 and these old
clusters is $2.4\pm 0.5\,$Gyr, showing that M68 is indeed a young GC
for its metallicity.  This statement could not be made using the \dv\
data and demonstrates the usefulness of using \dvb\ to probe relative
GC ages.  As high precision \dvb\ age estimates become available for
other metal-poor GCs, it will become possible to test the assumption
of CDKK that the 17 GCs in their sample had the same age.  

In their \dbv\ study, VandenBerg \ea\ (1990) determined that the
difference in colour between the turn-off and the base of the red
giant branch was the same for M68 and M92.  This implied that M68 had
the same age as M92 and other metal-poor GCs. This is somewhat
surprising, in light of the young age for M68 determined via \dvb.  To
explore this question further, we have performed a detailed comparison
of the CMDs for M92 and M68 using the deep M92 photometry of Stetson
\& Harris (1988), the M92 RR Lyrae photometry of Carney \ea\ (1991),
and the M68 photometry from Walker (1994). This comparison reveals
that, while VandenBerg \ea\ (1990) showed that M68 and M92 have nearly
identical \dbv\ values, the \dvb\ and \dv\ values of M68 differ by
approximately 0.1 magnitude as compared with those of M92.  Thus,
whereas \dbv\ appears to indicate that these two clusters have
identical ages (to within $\sim$0.3 Gyr), $\Delta$V shows that M68 is
$\sim$2 Gyr younger than M92. Salaris \ea\ (1993) and Zinn (1993) have
shown that, in general, ages derived via \dbv\ and \dv\ are in good
agreement.  However, Salaris \ea\ (1993) also point out that M68 is
one of a small number of clusters for which this is not the case, in
accordance with our result.  This implies that age is not the only
variable which differs between these two clusters.  It is not clear
what this other variable could be. It could simply be different
compositions (such as $[\alpha/{\rm Fe}]$), or it could be something
more exotic (such as rotation, or convection) which differs between
stars in these two clusters.  This is an interesting issue which we
are currently investigating.  Our ability to distinguish such small
differences between the $\Delta$V values of M68 and M92 is a testament
to the improvements made in the reduction and calibration of globular
cluster photometry and the use of the \dvb\ age diagnostic.

\section{Summary}\label{sec4}
An extensive Monte Carlo analysis indicates that the theoretical
uncertainty in \mvb\ is similar to \mvto.  The sensitivity of \mvb\ to
age changes is similar to \mvto.  The objective fitting technique
described in \S \ref{sec3} indicates that the error in measuring \vb\
in observational data is $\sim \pm 0.006\,$mag, at least an order of
magnitude smaller than the error typically quoted in V(TO).  Hence,
\mvb\ is a superior age indicator to \mvto.  We suggest that observers
should measure \vb\ as outlined in \S \ref{sec3}, and provide this value as a
routine part of the analysis of GC CMDs.  A calibration of age as a
function of \mvb\ (for B--V data) is presented in eq.\ 1 and
Table \ref{tab2}.  A similar calibration for V--I data is presented in
appendix \ref{app1}.

The use of \mvb\ as an age indicator requires a knowledge of the
distance modulus.  We suggest the use of the absolute magnitude of the
horizontal branch so that ages can be derived using the difference in
magnitude between the horizontal branch, and V(BTO), \dvb. This leads to
an age for M68 ($\feh = -2.1$) of $12.8\pm 0.3\,$Gyr, assuming $\mvrr
= 0.20\,\feh + 0.98$.  The error in the derived age is dominated by
the error in metallicity.  This is an internal error, useful in
comparing {\em relative} ages, and does not include the error due to
the uncertainty in the \mvrr\ calibration.  However, as shown by CDS,
many statements concerning relative ages are true independently of the
choice of \mvrr.  For example, M68 is significantly younger ($\sim
2.5\,$Gyr) than the mean age of 17 other low metallicity GC.  This
statement is true over the full range in \mvrr\ calibrations quoted in
the literature.  If the age of M68 had been determined using the
published value of V(TO), it would not be possible to state that M68
is a young GC, due to the large error in the derived age ($\pm
1.3\,$Gyr).  This demonstrates the unique advantage of using \mvb\ to
probe relative GC ages,  which should lead to new insights into the
formation of the Milky Way.

\section*{Acknowledgments}
L.M.K.\ was supported by the Department of Energy and funds provided by
Case Western Reserve University.  
A.S.\ was supported by NASA grant number HF-01077.01-94A from the
Space Telescope Science Institute, which is operated by the
Association of Universities for Research in Astronomy, Inc., under
NASA contract NAS5-26555.

\appendix
\section{Application to ($\rm V-I$) Data}\label{app1}
The discussion of this paper (and calibration presented in Table
\ref{tab2}) has centered on the use of ($\rm B-V$) data.  However the
use of \mvb\ to measure ages can be easily extended to other
colours.  Increasingly, observers have obtained GC CMDs using V, 
($\rm V-I$). The point \mvb\ may be defined as the magnitude of the point
which is brighter than the turn-off, and 0.05 mag redder in 
($\rm V-I$).  This will not correspond exactly to the same point defined in 
($\rm B-V$) data, but will be in a similar region of the HR diagram.
In order to facilitate the use of \mvb\ with ($\rm V-I$) data, Table
\ref{tab3} presents the fitting coefficients which can be combined
with eq.\ 1 to obtain the age of a GC, given \mvb in V, 
($\rm V-I$) data.
\begin{table}
\caption{Fit Coefficients for ($\rm V-I$) Data} \label{tab3}
\begin{tabular}{lrrr}
\hline\hline
\multicolumn{1}{c}{\feh}&
\multicolumn{1}{c}{$a_o$}&
\multicolumn{1}{c}{$a_1$}&
\multicolumn{1}{c}{$a_2$}\\[2pt]
\hline
$-2.5$& 1.971 & 0.1452 &$-0.002679$\\
$-2.0$& 2.148 & 0.1395 &$-0.002494$\\
$-1.5$& 2.228 & 0.1524 &$-0.002861$\\
$-1.0$& 2.462 & 0.1482 &$-0.002605$\\
$-0.5$& 2.578 & 0.1507 &$-0.002580$\\
\hline
\end{tabular}
\end{table}

\label{lastpage}

\begin{thebibliography}{}

\bibitem[\protect\citename{Ajhar et al.\ } 1996]{ajhar}Ajhar, E.A.\
et al.\  1996, AJ, 111, 1110


\bibitem[\protect\citename{Basu et al.\ } 1996]{basu} Basu, S.,
Christensen-Dalsgaard, J., Schou, J., Thompson, M.J. \& Tomczyk, S.,
1996, Ap. J., in press

\bibitem[\protect\citename{Bahcall } 1989]{bahcall}Bahcall, J.N.\
1989, Neutrino Astrophysics (Cambridge: Cambridge Univ. Press)

\bibitem[\protect\citename{Bahcall \& Pinsoneault }
1992]{bp92}Bahcall, J.N.\ \& Pinsonneault, M.H.\ 1992, Reviews of
Modern Physics, 64, 885

\bibitem[\protect\citename{Carney et al.\ } 1992]{carn}Carney, B. W.,
Storm, J., Trammell, S. R.\ \& Jones, R. V.\ 1992, PASP, 104, 44

\bibitem[\protect\citename{Chaboyer } 1995]{liege} Chaboyer, B.\
1995, in {\it Stellar Evolution: What Should Be Done?\/}, eds.\ A.\
Noels, D.\ Fraipont-Caro, M.\ Gabriel, N.\ Grevesse \& P.\ Demarque
(Li\`ege: Institut d'Astro\-physique), 345

\bibitem[\protect\citename{CDS}]{cds}Chaboyer, B., Demarque, P.\ \&
Sarajedini, A.\ 1996, ApJ, 459, 558

\bibitem[\protect\citename{Chaboyer et al.\ } 1996]{cdkk}Chaboyer,
B., Demarque, K., Kernan, P.J.\ \& Krauss, L.M.\ 1996, Science, 271,
957

\bibitem[\protect\citename{Chaboyer \& Kim } 1995]{yc}Chaboyer, B.\
\& Kim, Y.-C.\ 1995, ApJ, 454, 767

\bibitem[\protect\citename{Da Costa \& Armandroff } 1995]{da95}Da
Costa, G.S.\ \& Armandroff, T. 1995, AJ, 109, 2533

\bibitem[\protect\citename{Demarque, Guenther \& Kim
}1996a]{pie1}Demarque, P., Guenther, D.B.\ \& Kim, Y.-C.\ 1996a, ApJ,
submitted

\bibitem[\protect\citename{Demarque, Guenther \& Kim }
1996b]{pie2}Demarque, P., Guenther, D.B.\ \& Kim, Y.-C.\ 1996b, in
{\it Stellar Evolution: What Should Be Done?\/}, eds.\ A.\ Noels, D.\
Fraipont-Caro, M.\ Gabriel, N.\ Grevesse \& P.\ Demarque (Li\`ege:
Institut d'Astro\-physique), 279

\bibitem[\protect\citename{Green, Demarque \& King } 1987]{ryi}Green,
E.M., Demarque, P.\ \& King, C.R.\ 1987, The Revised Yale Isochrones
\& Luminosity Functions (New Haven: Yale Univ. Obs.)

\bibitem[\protect\citename{Guenther et al.\ } 1992]{guenth}Guenther,
D.B., Demarque, P., Kim, Y.-C.\ \& Pinsonneault, M.H.\ 1992, 387, 372

\bibitem[\protect\citename{Iglesias \& Rogers } 1991]{opal}Iglesias,
C.A.\ \& Rogers, F.J.\ 1991, ApJ, 371, 408

\bibitem[\protect\citename{Kim et al.\ } 1995]{kim1}Kim, Y.-C., Fox,
P.A., Sofia, S.\ \& Demarque, P.\ 1995, ApJ, 442, 422

\bibitem[\protect\citename{Kim et al.\ } 1996]{kim2}Kim, Y.-C., Fox,
P.A., Demarque, P.\ \& Sofia, S.\ 1996, ApJ, 461, 000 (in press)

\bibitem[\protect\citename{Krishna-Swamy } 1966]{ks}Krishna-Swamy,
K.S.\ 1966, ApJ, 145, 176

\bibitem[\protect\citename{Kurucz } 1991]{kur}Kurucz, R.L.\ 1991, in
Stellar Atmospheres: Beyond Classical Models, ed.\ L.\ Crivellari, I.\
Hubeny, D.G.\ Hummer, (Dordrecht: Kluwer), 440

\bibitem[\protect\citename{Kurucz } 1992]{kurcol} Kurucz, R.L.\ 1992,
in IAU Symp.\ 149, The Stellar Populations of Galaxies, ed.\ B.\
Barbuy, A.\ Renzini, (Dordrecht: Kluwer), 225

\bibitem[\protect\citename{Kraft } 1994]{kraft}Kraft, R.P.\ 1994,
PASP, 106, 553

\bibitem[\protect\citename{Langer et al.\ } 1983]{langer}Langer,
G.E., Kraft, R.P., Carbon, D.F., Friel, E.\ \& Oke, J.B.\ 1986, PASP,
98, 473

\bibitem[\protect\citename{Michaud \& Proffitt } 1993]{mp}Michaud,
G.\ \& Proffitt, C.R. 1993, in Inside the Stars, IAU Col.\ 137, ed.\
A.\ Baglin \& W.W.\ Weiss (San Fransico: PASP), 246

\bibitem[\protect\citename{Nissen et al.\ } 1994]{nissen}Nissen, P.,
Gustafsson, B., Edvardsson, B.\ \& Gilmore, G.\ 1994, A\&A, 285, 440

\bibitem[\protect\citename{Renzini } 1991]{renzini}Renzini, A.\ 1991,
in Observational Tests of Cosmological Inflation, eds.\ T.\ Shanks, et
al.\ , (Dordrecht: Kluwer), 131

\bibitem[\protect\citename{Salaris, Chieffi \& Straniero }
1993]{scs}Salaris, M.\ Chieffi,A.\ \& Straniero, O. 1993, ApJ, 414,
580

\bibitem[\protect\citename{Sarajedini \& Demarque }
1990]{cd}Sarajedini, A.\ \& Demarque, P.~1990, ApJ, 365, 219

\bibitem[\protect\citename{Sarajedini \& King }
1989]{sk}Sarajedini, A.\ \& King, C.R.\ 1989, AJ, 98, 1624


\bibitem[\protect\citename{Stetson \& Harris } 1988]{shar}Stetson,
P. B.\ \& Harris, W. E.\ 1988, AJ, 96, 909

\bibitem[\protect\citename{van Altena, Lee \& Hoffleit }
1995]{ypc}van Altena, W.A., Lee, John Truan-liang, \& Hoffleit, D.\
1995, The General Catalogue of Trigonometric Parallaxes (Yale
University Observatory: New Haven, CT)

\bibitem[\protect\citename{VandenBerg, Bolte \& Stetson } 1990]{vbs}
VandenBerg, D.A., Bolte, M.\ \& Stetson, P.B.\ 1990, AJ, 100, 445

\bibitem[\protect\citename{Walker } 1994]{walker}Walker, A.R.\ 1994,
AJ, 108, 555

\bibitem[\protect\citename{Zinn } 1993]{zinn}Zinn, R.\ 1993, in The
Globular Cluster-Galaxy Connection, eds. G.H.\ Smith \& J.P.\ Brodie
(San Francisco: ASP), 38

\bibitem[\protect\citename{Zinn \& West } 1984]{zinnw}Zinn, R.\ \&
West, M.\ 1984, ApJS, 55, 45

\end{thebibliography}
\end{document}